%%%%%%%%%%%%%%%%%%%%%%%%%%%%%%%%%%%%%%%%%%%%%%%%%%%%%%%%%%%%%%%%%%%%%%
\input harvmac
%--------------------------------------------------
% personal macros
%---------------------------------------------------
\def\a{\alpha}    \def\b{\beta}              
     \def\f{\phi}       
\def\g{\gamma}               
\def\L{\Lambda}                    \def\r{\rho}
        \def\O{\Omega}     
                 
\def\t{\tau}        \def\w{\varphi}    

%------------------------------------------------
% Euler Fonts
\font\teneufm=eufm10
\font\seveneufm=eufm7
\font\fiveeufm=eufm5
\newfam\eufmfam
\textfont\eufmfam=\teneufm
\scriptfont\eufmfam=\seveneufm
\scriptscriptfont\eufmfam=\fiveeufm
\def\eufm#1{{\fam\eufmfam\relax#1}}
\font\teneusm=eusm10
\font\seveneusm=eusm7
\font\fiveeusm=eusm5
\newfam\eusmfam
\textfont\eusmfam=\teneusm
\scriptfont\eusmfam=\seveneusm
\scriptscriptfont\eusmfam=\fiveeusm

\font\tenmsx=msam10
\font\sevenmsx=msam7
\font\fivemsx=msam5
\font\tenmsy=msbm10
\font\sevenmsy=msbm7
\font\fivemsy=msbm5
\newfam\msafam
\newfam\msbfam
\textfont\msafam=\tenmsx  \scriptfont\msafam=\sevenmsx
  \scriptscriptfont\msafam=\fivemsx
\textfont\msbfam=\tenmsy  \scriptfont\msbfam=\sevenmsy
  \scriptscriptfont\msbfam=\fivemsy

%\def\eufm#1{{\bf #1}}  %If you do not have AMSfonts 2.1,
%\def\eusm#1{{\cal #1}} %uncomment these three lines
%\def\msbm#1{{\bf #1}}  %and comment the lines below %Euler Fonts
%-----------------------------------------------------
%Math

\def\fr#1#2{{\textstyle{#1\over#2}}}

\def\roughly#1{\raise.3ex\hbox{$#1$\kern-.75em\lower1ex\hbox{$\sim$}}}

%
%Journals

\def\pl#1#2#3{Phys.\ Lett.\ {{\bf #1}} {(#2)} {#3}}
\def\np#1#2#3{Nucl.\ Phys.\ {{\bf #1}} {(#2)} {#3}}
\def\prd#1#2#3{Phys.\ Rev.\ {{\bf #1}} {(#2)} {#3}}
\def\prl#1#2#3{Phys.\ Rev.\ Lett.\ {{\bf #1}} {(#2)} {#3}}

%--------------------------------------------------------
%Macros specific to this file

\def\hS{{\widehat S}}

\def\cha{\cosh\alpha}
\def\chb{\cosh\beta}
\def\chg{\cosh\gamma}
\def\sha{\sinh\alpha}
\def\shb{\sinh\beta}
\def\shg{\sinh\gamma}

\def\pr{\prime}

\def\gBE{\eufm{g}_{\raise-.1ex\hbox{${}_\BE$}}}
\def\gEc{\eufm{g}_{\raise-.1ex\hbox{${}_E$}}^C}

%---------------------------------------------------------
\lref\SV{
A.\ Strominger and C.\ Vafa, 
\pl{B 379}{1996}{99}.
}
\lref\Pol{
J.\ Polchinski, 
{\it TASI Lectures on D-branes},
hep-th/9611050 and references therein.
}
\lref\MS{
J.\ Maldacena and A.\ Strominger, 
\prl{77}{1996}{428}.
}
\lref\JKM{
C. Johnson, R. Khuri, and R. Myers, 
\pl{B 378}{1996}{78}.
}
\lref\BMPV{
J. C. Breckenridge, R. C. Myers, A. W. Peet, and C. Vafa, 
{\it D-branes and Spinning Black Holes},
hep-th/9602065:
J. C. Breckenridge, D. A. Lowe, R. C. Myers, A. W. Peet, 
A. Strominger, and C. Vafa 
\pl{B 381}{1996}{423}.
}
\lref\CM{
C.\ G.\ Callan and J.\ M.\ Maldacena,
\np{B 472}{1996}{591}.
}
\lref\HS{
G. T. Horowitz and A. Strominger,
\prl{77}{1996}{2368}.
}
\lref\DMW{
A. Dhar, G. Mandal, and S. R. Wadia,
\pl{B 388}{1996}{51}.
}
\lref\DMa{
S. R. Das and S. D. Mathur,
\np{B 478}{1996}{561}:
\np{B 482}{1996}{153}.
}
\lref\GKa{
S. Gubser and I. Klebanov,
\np{B 482}{1996}{173}.
}
\lref\MSb{
J.\ Maldacena and A.\ Strominger, 
\prd{D 55}{1997}{861}.
}
\lref\GKb{
S. Gubser and I. Klebanov,
\prl{77}{1996}{4491}.
}
\lref\DGM{
S. R. Das, G. Gibbons, and S. D. Mathur,
{\it Universality of Low Energy Absorption Cross Sections for
Black Holes},
hep-th/9609052.
}
\lref\MSc{
J.\ Maldacena and A.\ Strominger, 
{\it Universal Low-Energy Dynamics for Rotating Black Holes},
hep-th/9702015.
}
\lref\MNY{
M.\ D.\ McGuigan, C.\ R.\ Nappi, and S.\ A.\ Yost,  
\np{B 375}{1992}{421}.
}
\lref\BTZ{
M.\ Banados, C.\ Teitleboim, and J.\ Zanelli,
\prl{69}{1992}{1849}.
}
\lref\HW{
G. T. Horowitz and D. L. Welch,
\prl{71}{1993}{328}.
}
\lref\BHO{
E.\ Bergshoeff, C.\ Hull, and T.\ Ortin,
\np{B 451}{1995}{547}
}

%--------------------------------------------------------

\baselineskip=15pt plus 1.2pt minus .6pt
\newskip\normalparskip
\normalparskip = 8pt plus 1.2pt minus .6pt
\parskip = \normalparskip
\parindent=18pt
\def\submit{\baselineskip=20pt plus 2pt minus 2pt}
\def\ack{\bigbreak\bigskip\bigskip\centerline{{\bf
Acknowledgements}}\nobreak}
%--------------------------------------------------------
\font\Titlerm=cmr10 scaled\magstep2
\nopagenumbers
\rightline{hep-th/9704005}
\vskip .4in
\centerline{\fam0\Titlerm U-duality between Three and
Higher Dimensional Black Holes}
\tenpoint\vskip .4in\pageno=0%}

\centerline{
Seungjoon Hyun
}
\centerline{{\it Department of Physics and 
Research Institute for Basic Sciences }}
\centerline{{\it  Kyunghee University}}
\centerline{{\it Seoul 130-701, Korea}}
\centerline{{\it (sjhyun@nms.kyunghee.ac.kr)}}
\vskip .3in

\noindent
\abstractfont
We show that the D-brane configurations for the  five and
four-dimensional  black holes
give the geometry of 
two and three-dimensional ones as well.
 The emergence of these lower dimensional black holes from the 
D-brane configurations for those of higher dimensions
comes from the choice of the integration constant of 
harmonic functions, which decides the asymptotic behavior of 
the metric and other fields. We show that 
they are equivalent, which are connected by U-dual transformations.
This means that stringy black holes in various dimensions
are effectively in the same universality class and many properties
of black holes in the same class can be infered from the study
of those of the three-dimensional black holes.

\tenpoint

\Date{March, 1997}
%\draftmode
%\baselineskip=14pt plus 1.2pt minus .6pt
\submit

Recent developments \SV\ in nonperturbative aspects of string theory,
especially 
D-branes \Pol, made it possible to give a statistical interpretation
on the black hole entropy by 
counting microscopic states.
The entropy from the degeneracies of BPS-saturated D-brane
bound states exactly agrees with the Bekenstein-Hawking entropy
of the corresponding
five \SV\ and four-dimensional extremal black holes \MS\JKM\
and for spinning ones \BMPV\ and the near-extremal cases \CM\HS.
Remarkably, the calculations of other quantities like
decay and emission rates \CM\DMW\DMa\GKa\ and
grey body factors \MSb\GKb\ of near-extremal D-brane configurations
also exactly agree with those from the semiclassical Hawking radiation
of the corresponding black holes.
%See also \others\
All these results indicate that there seems to be universal
behavior \DGM\ among various black holes and many essential features
of low energy effective theory can be read from semiclassical
analysis of the general black
hole solution of ordinary Einstein-Maxwell gravity, without
using string theory \MSc.

In this letter, we show that the same D-brane configurations 
as those of five and four-dimensional black holes also give
the geometry of those in two and three dimensions.
The configurations of the BPS states in string theory,
which preserves part of the supersymmetries,
are determined by the set of harmonic functions
$H_i(r)$ of transverse coordinates.
The arbitrary integration constants,
$H_i(\infty)$ of those functions are usually chosen to
be 1 to have the asymptotically Minkowskian metric.
Quite naturally, by choosing different value for $H_i(\infty)$,
namely $H_i(\infty)=0$ for some $i$, one would get different
background geometry, with different asymptotic behavior and singularity
structures, yet the same configurations would garantee to give
the same physics
from the string point of view. We show this explicitly, 
by finding U-dual transformations which connect those solutions.

We begin with type IIB theory on $T^4\times S^1$ and configurations of 
$Q_1$ fundamental strings wrapping $S^1$ and 
$Q_5$ solitonic five-branes wrapping $T^4\times S^1$, along
with the momentum on $S^1$. They are the
S-dual version of the configurations of D-strings, D5-branes and
momentum, used in the calculation of the entropy
of five-dimensional black holes.
The metric, dilaton and the non-vanishing components of
antisymmetric second rank tensor take the form
\eqn\sola{
\eqalign{
ds^2 = 
&H_1^{-1}(-dt^2+dx^2+K(dt-dx)^2)+
H_5(dx_1^2+\cdots +dx_4^2)+ dx_5^2 + \cdots + dx_8^2 ,
\cr
e^{-2\f}=&H_5^{-1} H_1 ,\cr
B_{tx}=&H_1^{-1}-1,\cr
H_{ijk}=&(dB)_{ijk}=\fr{1}{2}\epsilon_{ijkl}\partial_{l}H_5,
\ \ \ i,j,k,l=1,2,3,4,\cr
}
}
where $H_1$, $H_5$ and $K$ are harmonic functions of 
$x_1, \cdots, x_4$. As is well-known, this configuration 
preserves $\fr{1}{8}$ supersymmetries.
It also has U-duality invariance %\HT\Sena\Senb\ 
under which $(H_1, H_5, K+1)$ permutes.
The five-dimensional black holes correspond
to the usual ansatz
\eqn\harma{
\eqalign{
H_1=1+\fr{r_1^2}{r^2},\ \ \ 
H_5=1+\fr{r_5^2}{r^2},\ \ \ 
K=\fr{r_k^2}{r^2}.
}
}

As noted above, we can use different harmonic functions which 
might give the metric describing different background geometry.
If we take those harmonic functions of the form
\eqn\harmb{
\eqalign{
H_1=\fr{r_1^2}{r^2},\ \ \
H_5=\fr{r_5^2}{r^2},\ \ \
K=\fr{r_k^2}{r^2},
}
}
then, after the dimensional reduction on $T^4$,
the six-dimensional metric becomes
\eqn\solb{
\eqalign{
ds^2 =& \fr{r^2}{r_1^2}(-dt^2+dx^2+K(dt-dx)^2)
+\fr{r_5^2}{r^2}dr^2+r_5^2d\O_3^2.
\cr
}
}
Now $ d\O_3^2$ part is completely decoupled, and 
the geometry turns out to be  $(AdS)_3\times S^3\times T^4$, where
$(AdS)_3$ denotes three-dimensional extremal
anti-de Sitter black holes given in \BTZ.
Note that if the radius $r_5$ of $S^3$  is small,
the geometry can be interepreted as
the spontaneous compactification down to three dimensions by
$H=dB=-r_5^2\epsilon_{3}$,
where $\epsilon_{3}$ is the unit volume element of $S^3$.
After replacing by $r^2+r_k^2 \longrightarrow r^2$ and
$\fr{x}{\sqrt{r_1r_5}}\longrightarrow \w$,
the Einstein metric is shown to describe the
three-dimensional extremal black holes, tensored by $S^3$,
\eqn\solbb{
\eqalign{
ds_E^2 =& - \fr{(r^2-r_0^2)^2}{l^2r^2}dt^2
+r^2(d\w-\fr{r_0^2}{lr^2}dt)^2
+\fr{l^2r^2}{(r^2-r_0^2)^2}dr^2
+l^2d\O_3^2.
\cr
}
}
Here $l=\sqrt{r_1r_5}$ is the radius of compactified space $S^3$,
which is related to effective three-dimensional cosmological 
constant $\L=-l^{-2}$, and $r_k$ becomes the radius of event
horizon. 
Note that the non-vanishing entropy of this three-dimensional
extremal black hole can be easily deduced  from D-brane calculations.

Another choice for the harmonic functions of the form
\eqn\harmc{
\eqalign{
H_1=1+\fr{r_1^2}{r^2},\ \ \ 
H_5=\fr{r_5^2}{r^2},\ \ \ 
K=\fr{r_k^2}{r^2}.
}
}
generates effective two-dimensional charged black hole solutions.
Under the dimensional reduction on $T^4\times S^1$ and the same
spontaneous compactification on $S^3$ as above, the solution becomes
\eqn\solc{
\eqalign{
ds^2 =& -(1+\fr{r_1^2}{r^2})^{-1}(1+\fr{r_k^2}{r^2})^{-1}dt^2
+\fr{r_5^2}{r^2}dr^2,
\cr
e^{-2\f}=&r_5^{-2}(r^2+r_1^2)^{\fr{1}{2}}
(r^2+r_k^2)^{\fr{1}{2}},\cr
}
}
where the case $r_1=r_k$ corresponds to the extremal limit of
two-dimensional charged black holes studied in \MNY.

So far, we have shown that completely different background
geometries emergy from the same D-brane configurations, depending
on the choice of the integration constant.
Now we show that 
they are related by $U$-duality, thus indeed equivalent from the string
point of view.
Starting with \sola-\harma\ (dropping the irrelevant $T^4$ part),
we change the coordinates as follows:
\eqn\coorda{
\eqalign{  
t\longrightarrow 2t-\fr{1}{2}x ,\ \ \
x\longrightarrow\fr{1}{2}x ,
}
}
in which the fields are given by
\eqn\sold{
\eqalign{
ds^2 = 
&-\fr{1}{KH_1}dt^2+\fr{K}{H_1}(dx+\fr{1-2K}{K}dt)^2
+H_5(dx_1^2+\cdots +dx_4^2),
\cr
e^{-2\f}=&H_5^{-1} H_1 ,\cr
B_{tx}=&H_1^{-1}-1,\cr
H_{ijk}=&(dB)_{ijk}=\fr{1}{2}\epsilon_{ijkl}\partial_{l}H_5,
\ \ \ i,j,k,l=1,2,3,4,\cr
}
}
After taking the T-dual transformation \BHO\
with respect to $x$-coordinate, we get
\eqn\sole{
\eqalign{
ds^2 = 
&-\fr{1}{KH_1}dt^2+\fr{H_1}{K}(dx+\fr{1-H_1}{H_1}dt)^2
+H_5(dx_1^2+\cdots +dx_4^2),
\cr
e^{-2\f}=&\fr{K}{H_5} ,\cr
B_{tx}=&(K)^{-1}-1,\cr
H_{ijk}=&(dB)_{ijk}=\fr{1}{2}\epsilon_{ijkl}\partial_{l}H_5,
\ \ \ i,j,k,l=1,2,3,4,\cr
}
}
where we made constant gauge transformation for $B_{tx}$ by adding 1.
This T-dual transformation
maps from one background geometry to another, which are related by
$H_1^\pr=K$, $H_5^\pr=H_5$, and $K^\pr=H_1-1$.
It turns the five-dimensional black
hole solutions with \harma\ into effective two-dimensional
black hole solutions \solc.
By performing U-dual transformations with
$U=ST_5T_6T_7T_8S$, where $S$ denotes the S-dual transformation in
ten-dimensional type IIB 
theory and $T_i$ denotes the T-dual transformation on $i$-th coordinate,
we can transform ($H_1^\pr, H_5^\pr,K^\pr$)
to ($H_5^\pr, H_1^\pr,K^\pr$).
Now we apply the same
T-dual transformation, along with the same replacement as \coorda,
and get  $(H_1^{\pr\pr},H_5^{\pr\pr},K^{\pr\pr})=(H_1-1,K,H_5-1)$,
which corresponds to the three-dimensional
black hole solutions \solbb.

Now we turn to the D-brane configurations in type IIA theory
on $T^4\times S^1\times \hS^1$, which give the geometry of
the four-dimensional
extremal black holes \MS\JKM . It consists of
$Q_2$ D2-branes wrapping $S^1\times \hS^1$,
$Q_6$ D6-branes wrapping $T^4\times S^1\times \hS^1$,
$Q_5$ NS5-branes wrapping $T^4\times S^1$ and
the momentum on $S^1$. The metric and other fields have the form
\eqn\solf{
\eqalign{
ds^2 = 
&(H_2 H_6)^{-\fr{1}{2}}(-dt^2+dx^2+K(dt-dx)^2)+
H_5(H_2 H_6)^{-\fr{1}{2}}dx_4^2
\cr
+& H_2^{\fr{1}{2}} H_6^{-\fr{1}{2}}(dx_5^2 + \cdots + dx_8^2)
+H_5(H_2 H_6)^{\fr{1}{2}}(dx_1^2 + \cdots + dx_3^2),
\cr
e^{-2\f}=&H_5^{-1} H_2^{-\fr{1}{2}}H_6^{\fr{3}{2}} ,\cr
H_{ij4}=&\fr{1}{2}\epsilon_{ijk}\partial_{k}H_5,
\ \ \ i,j,k=1,2,3,\cr
C_{t4x}=&H_2^{-1}-1,\cr
(dA)_{ij}=&\fr{1}{2}\epsilon_{ijk}\partial_{k}H_6,
\ \ \ i,j,k=1,2,3.\cr
}
}          
The four-dimensional extremal black hole solutions correspond
to the harmonic functions of the form
\eqn\harmd{
\eqalign{
H_2=1+\fr{r_2}{r},\ \ \ 
H_5=1+\fr{r_5}{r},\ \ \  
H_6=1+\fr{r_6}{r},\ \ \ 
K=\fr{r_k}{r}.
}
}

If, instead, we choose 
\eqn\harmdb{
\eqalign{
H_2=\fr{r_2}{r},\ \ \ 
H_5=\fr{r_5}{r},\ \ \  
H_6=\fr{r_6}{r},\ \ \ 
K=\fr{r_k}{r},
}
}
we come up with the three-dimensional extremal black hole solutions,
with $S^2$ factor,
\eqn\solg{
\eqalign{
ds_E^2 =& - \fr{(\r^2-\r_0^2)^2}{l^2\r^2}d\t^2
+\r^2(d\w-\fr{\r_0^2}{l\r^2}dt)^2
+\fr{l^2\r^2}{(\r^2-\r_0^2)^2}d\r^2
+\fr{l^2}{4}d\O_2^2,
\cr
}
}
where $l=2(r_2r_5r_6)^{\fr{1}{3}}$,
$\t=2t$, $\r^2= \fr{l}{2}(r+r_k)$, $\w= \fr{2x}{l}$ and
$\r_0^2= \fr{l}{2}r_k$.
Again, it  can be shown to be U-dual to the four-dimensional solutions
\sole-\harmd.
Firstly, we change the coordinates $t$ and $x$ as \coorda\ and
perform U-dual 
transformations with $U=T_4T_5T_6T_7T_8ST_x$, which give the 
original configurations with new harmonic functions,
$(H_2^\pr, H_6^\pr, H_5^\pr, K^\pr)=(H_5, K, H_6, H_2-1)$.
By following  the same procedure, 
i.e. replacing by $t\longrightarrow 2t-\fr{1}{2}x_4$ and
$x_4\longrightarrow\fr{1}{2}x_4$ and
taking another U-dual
transformations with $U=T_xT_5T_6T_7T_8ST_4$, 
we get 
$$(H_2^{\pr\pr}, H_6^{\pr\pr}, H_5^{\pr\pr}, K^{\pr\pr})=
(H_5^\pr, K^\pr, H_6^\pr, H_2^\pr-1)=(H_6, H_2-1, K, H_5-1).
$$ 
Another change of coordinates, $t$ and $x$, as \coorda\ and U-dual
transformations with $U=T_4T_5T_6T_7T_8ST_x$  generate 
the three-dimensional black hole solution \solg\ with
$$(H_2^{\pr\pr\pr}, H_6^{\pr\pr\pr}, H_5^{\pr\pr\pr}, K^{\pr\pr\pr})=
(H_5^{\pr\pr}, K^{\pr\pr}, H_6^{\pr\pr}, H_2^{\pr\pr}-1)
=(K, H_5-1, H_2-1, H_6-1).
$$  

This equivalence can be easily extended to non-extremal black holes
as well. In the followings, as an example, we show the equivalence of
five and three-dimensional non-extremal black holes. Four-dimensional
case can be treated along the same way.
We begin with the configurations \CM\HS\
\eqn\solh{
\eqalign{
ds^2
=&-(f_1 f_k)^{-1}(1-\fr{r_0^2}{r^2})dt^2
+f_1^{-1}f_k(dx-\fr{r_0^2\chg \shg}{r^2f_k} dt)^2\cr
+&f_5((1-\fr{r_0^2}{r^2})^{-1}dr^2+r^2 d\O_3^2)+
dx_5^2 + \cdots + dx_8^2,
\cr
e^{-2\f}=&f_5^{-1} f_1 ,\cr
B_{tx}=&\fr{\cha}{\sha}(f_1^{-1}-1),\cr
H_{ijk}=&(dB)_{ijk}=\fr{\chb}{2\shb}\epsilon_{ijkl}\partial_{l}f_5,
\ \ \ i,j,k,l=1,2,3,4,\cr
}
}
where $f_1$, $f_5$ and $f_k$ are given by
\eqn\harme{
\eqalign{
f_1=1+\fr{r_0^2\sinh^2\a}{r^2},\ \ \
f_5=1+\fr{r_0^2\sinh^2\b}{r^2},\ \ \
f_k=1+\fr{r_0^2\sinh^2\g}{r^2}.
}
}
After changing the coordinates as
\eqn\coordb{
\eqalign{  
t\longrightarrow(\fr{\chg+\shg}{\shg})^2t-
(\fr{\shg}{\chg+\shg})^2x ,\ \ \
x\longrightarrow(\fr{\shg}{\chg+\shg})^2x ,
}
}
and taking the T-dual transformation with respect to $x$,
the metric and other fields become
\eqn\soli{
\eqalign{
ds^2  =&-(h_1 h_k)^{-1}(1-\fr{r_0^2}{r^2})dt^2
+h_1^{-1}h_k(dx-\fr{r_0^2\cha \sha}{r^2h_k} dt)^2\cr
+&f_5((1-\fr{r_0^2}{r^2})^{-1}dr^2+r^2 d\O^2)+ dx_5^2 + \cdots + dx_8^2,
\cr
e^{-2\f}=&f_5^{-1} h_1 ,\cr
B_{tx}=&(h_1^{-1}-1),\cr
H_{ijk}=&(dB)_{ijk}=\fr{\chb}{2\shb}\epsilon_{ijkl}\partial_{l}f_5,
\ \ \ i,j,k,l=1,2,3,4,\cr
}
}
where
\eqn\harmf{
\eqalign{
h_1=f_k-1=\fr{r_0^2\sinh^2\g}{r^2},\ \ \
h_k=f_1  =1+\fr{r_0^2\sinh^2\a}{r^2}.
}
}
As is the case for extremal limit, the metric describes two-dimensional
charged black hole solutions which reduce to the solutions in \MNY\
under the limit $\a=\g$.
After we take the same U-dual transformations as the extremal case
and perform the similar steps as above,
we finally get \solh\ with
\eqn\harme{
\eqalign{
f_1=\fr{r_0^2\sinh^2\a}{r^2},\ \ \
f_5=\fr{r_0^2\sinh^2\b}{r^2},\ \ \
f_k=1+\fr{r_0^2\sinh^2\g}{r^2}.
}
}
The corresponding Einstein metric describes
the geometry of three-dimensional black holes in \BTZ, with
trivial $S^3\times T^4$ factor, as follows:
\eqn\solj{
\eqalign{
ds_E^2 =& - \fr{(\r^2-\r_+^2)(\r^2-\r_-^2)}{l^2\r^2}dt^2
+r^2(d\w-\fr{\r_+\r_-}{\r^2}dt)^2
+\fr{l^2\r^2}{(\r^2-\r_+^2)(\r^2-\r_-^2)}d\r^2 +l^2d\O_3^2, \cr
}
}
where
$\r^2= r^2+r_0^2\sinh^2\g$, $\r_+^2= r_0^2\cosh^2\g$,
$\r_-^2= r_0^2\sinh^2\g$,
$l^2=r_0^2\sha\shb$ and $\w= \fr{x}{l}$.

In this paper, we have shown that, as far as strings are concerned,
four and five-dimensional black holes, which have four and three
charges, respectively, are U-dual, and thus
equivalent, to the effective three-dimensional
black holes, even though, in contrast to higher dimensional black
hole solutions, those of three dimensions
have constant dilaton and, most notably, has no curvature
singularities. This kind of phenomenon  is not
new in the string theory \HW.

 From the string point of view, there may be only a few universality
classes, whose representatives would be, hopefully, lower
dimensional black holes.
If this is the case, it helps us understanding the universal
behavior among various black holes.
Futhermore, from the ordinary field theory point of view, treating
gravitational systems in these lower dimensions is much easier 
than those of higher dimensions, hence it may be possible
to get deeper insights underlying connections between the D-brane
calculation and ordinary semiclassical approach.
The detailed study on the
effective three-dimensional black holes will be reported elsewhere.

\ack{  
This work was supported in part by Korea Research
Foundation. }

\listrefs
\bye